\documentclass [12pt]{article}
\usepackage{amsmath,amssymb,epsfig,cite}

\numberwithin{equation}{section}

\begin{document}

\begin{center}
{\bf FATE OF THE SUPERCONDUCTING GROUND STATE ON THE MOYAL PLANE}
\bigskip

Prasad Basu$^a$\footnote{prasad@cts.iisc.ernet.in},
Biswajit Chakraborty$^{b,c}$\footnote{biswajit@bose.res.in} and 
Sachindeo Vaidya$^a$\footnote{vaidya@cts.iisc.ernet.in} \\
\smallskip
$^{a}${\it Centre for High Energy Physics, Indian Institute of Science, Bangalore 560 012, India} \\
$^b${\it S. N. Bose Centre for Basic Sciences, Block JD Sector III, Saltlake, Kolkata 700098 India}\\
$^c${\it Centro Brasileiro de Pesquisas Fisicas (CBPF) Rua Dr.Xavier Sigaud 150, cep 22290-180, Rio de Janeiro, Brazil}

\end{center}

\begin{abstract}
It is known that Berry curvature of the band structure of certain crystals can lead to effective 
noncommutativity between spatial coordinates. Using the techniques of twisted quantum field theory, 
we investigate the question of the formation of a paired state of twisted fermions in such a system. 
We find that to leading order in the noncommutativity parameter, the gap between the non-interacting 
ground state and the paired state is {\it smaller} compared to its commutative counterpart. This 
suggests that BCS type superconductivity, if present in such systems, is more fragile and easier to 
disrupt.
\end{abstract}

\vskip 1.0 cm
\section{Introduction}

Noncommutative quantum theories have the potential to provide us with
insights not only at Planck scale physics (see for example \cite{dop})
but also in the domain of condensed matter and statistical physics
(see \cite{ML,Sch1,Sch2} for a recent reviews). Indeed, apart from the
well known noncommutativity of the guiding centre coordinates in the
Landau problem, it has been shown by Xiao et. al.\cite{Xiao} that due
to Berry curvature of the band structure, the quantum mechanics for
the electrons in certain materials is govened by effective
noncommutativity between phase space variables. Specifically, this
noncommutativity takes the form
\begin{eqnarray}
\bigl[\hat{x}_{i},\hat{x}_{j}\bigr ]&=&\frac{i\Omega \epsilon_{ij}}{1+(e/\hbar)B\Omega}, \nonumber\\ 
\bigl[\hat{k}_i, \hat k_j\bigr]&=&-\frac{i(e/\hbar)B
  \epsilon_{ij}}{1+(e/\hbar)B\Omega}, \quad \quad \hat p_i \equiv \hbar \hat k_i\nonumber \\
\bigl[\hat x_{i},\hat k_j\bigr]&=& \frac{i \delta_{ij}}{1+(e/\hbar)B\Omega} 
\end{eqnarray} 
where $i,j=1,2$ and $B$ is the external magnetic field (along the
third direction) and $\Omega$ the Berry curvature arising due to the
breaking of time reversal symmetry (for example in certain
ferromagnetic materials) or spatial inversion symmetry (for example in
GaAs). Although the possibility of superconductivity in such materials has
already received some attention in the literature (see for
e.g. \cite{sigr,cond06}), our treatment of this question will use
techniques arising from quantum fields defined on noncommutative spacetimes.

In the absence of external magnetic field ($B=0$) these
commutation relations take the form
\begin{eqnarray}
\bigl[\hat x_{i},\hat x_{j}\bigr ]&=&i\Omega, \nonumber \\ 
\bigl[\hat k_i, \hat k_j\bigr]&=&0, \nonumber \\
\bigl[\hat x_{i},\hat k_j\bigr]&=& i \delta_{ij}, \quad i,j=1,2.
\end{eqnarray} 
Quantum field theories on such a noncommutative space (also known as
the Moyal plane) admit a novel quantization called {\it twisted
  quantization} \cite{bmpv,bgmpqv} that implements the underlying
spacetime symmetries using a new coproduct, modifying the canonical
(anti)commutation relations of the basic oscillator algebra for
creation and annihilation operators. Such deformed (anti) commutation
relations can have dramatic consequences, like say the violation of
the Pauli principle \cite{Sch1,bpqv} at length scales of order
$\sqrt{\Omega}$ and the energy shift in the model of degenerate
electron gas \cite{dgen}.

In this article, we shall address the question of the formation of
paired bound states (even parity spin singlet states of standard BCS
theory) of twisted fermions on the Moyal plane. The corresponding
question in ordinary space was addressed long ago by Cooper, who
showed that formation of such bound states is indeed possible in the
presence of arbitrarily weak attractive interaction between fermions
of opposite momenta and spins. This makes the
Fermi surface unstable, resulting the formation of BCS state which is
a coherent state of Cooper pairs.  The existence of BCS state explains
various key features of superconductivity.

In the next section, we begin by revisiting the usual calculation (due
to Cooper) in the second quantized formalism. The discussion is
standard textbook material, and may be found for example, in
\cite{Plichke}.  We then extend it to the Moyal plane in section 3 and
derive the new gap equation.  This equation can be solved for small
values of the parameter $\Omega$, and we find that the presence of
$\Omega$ leads to a reduction the the gap, suggesting that it is
comparatively easier to disrupt the superconducting state. We conclude
in section 4 with a brief discussion of the implications of our
results.

\section{Commutative Case}
Let us consider a many-fermion state in which there are additional two fermions above the filled Fermi 
sea.  These two fermions have equal and opposite momenta and spins, and their energies are 
(slightly) higher than the Fermi energy. In the second quantized formalism the quantum state 
representing this system is
\begin{eqnarray} 
| \psi_{\vec k}\rangle &=& c_{\vec k, \frac{1}{2}}^\dagger c_{-\vec k,
  -\frac{1}{2}}^\dagger |F\rangle \quad k > k_{F}, \quad {\rm with}\label{pair} \\
|F\rangle &=& \prod_{k\leq k_{F},\sigma}c^{\dagger}_{\vec k
  ,\sigma}|0\rangle, \quad \sigma = \pm \frac{1}{2} \label{2}
\end{eqnarray} 
where $ c^{\dagger}_{\vec k,\sigma}$ is the creation operator of an fermion of momentum $\vec k$ 
and spin $\frac{\hbar}{2}\sigma$. A generic two-particle state $|\psi \rangle$ can be written as 
\begin{equation} 
|\psi \rangle = \int d^3 k \,g(\vec{k}) |\psi_{\vec{k}}\rangle
\end{equation} 
The effective Hamiltonian describing the dynamics of the fermion pair is given by
\begin{eqnarray} 
\hat H &=& \hat H^0+\hat H^{int}, \quad {\rm with} \label{hamBCS1}  \\
\hat H^0 &=& \int_{k_F} d^3k \, \epsilon(\vec k) c^{\dagger}_{\vec k, \sigma}
c_{\vec k, \sigma}  \label{hamBCS2}
\end{eqnarray}
and $\epsilon(\vec k)=\frac{\vec{k}^2}{2m}$ is energy of a single free fermion. The interaction part of 
the Hamiltonian in the position space representation $\hat H^{int}$ is
\begin{eqnarray}
\hat H^{int} = \frac{1}{2}\sum_{\sigma , \sigma'}\int d^3x \int d^3y \,
\psi_{\sigma}^{\dagger}(\vec x) \psi_{\sigma'}^{\dagger}(\vec y)
\psi_{\sigma'}(\vec y)\psi_{\sigma}(\vec x) v(\vec x,\vec y) 
\label{intH1}
\end{eqnarray}
where $v(\vec x,\vec y)=v(\vec x -\vec y)$ is the interaction potential between the fermion pair. Writing 
\begin{eqnarray}
\psi_{\sigma}(\vec x)=\int d^3k \, e^{-i\vec k.\vec x}c_{\vec
  k,\sigma} \quad {\rm and} 
\quad \tilde V(\vec k)=\int d^3x \, v(\vec x)e^{i\vec k.\vec x},
\label{psiandv}
\end{eqnarray}
the interaction Hamiltonian $\hat H^{int}$ takes the form
\begin{equation}
\hat H^{int} = \frac{1}{2}\sum_{\sigma, \sigma'}\int d^3p \,d^3q \,d^3r\,\tilde V(\vec r
-\vec q) c^{\dagger}_{\vec p,\sigma} c^{\dagger}_{\vec q,\sigma'}
c_{\vec r,\sigma'} c_{(\vec p+ \vec q -\vec r),\sigma}
\label{intH2}
\end{equation}
Solving the eigenvalue equation
\begin{equation}
\hat H |\psi\rangle=E|\psi\rangle
\label{Seqn1}
\end{equation}
for a generic two-particle state $|\psi\rangle$, we get
\begin{equation}
[2\epsilon(\vec k)-E]=-\int d^3k' \,\tilde V(\vec k-\vec k')g(\vec k').
\label{Seqn2}
\end{equation}
Choosing $\tilde V(\vec k-\vec k')$ to be a constant $-V$ for $k_F\leq k,k'\leq k_F +k_D$ and 
zero otherwise, and setting $\hbar^2 k_D^2/2m=\hbar \omega_D$ (where $\omega_D$ is the Debye
frequency) we can rewrite (\ref{Seqn2}) as
\begin{equation}
[2\epsilon_{k}-E]g(\vec k)= V\int d^3k' \, g(\vec k'), \quad k_F\leq k,k'
\leq k_F + k_D
\label{selfconsistent1}
\end{equation}
The RHS of (\ref{selfconsistent1}) is independent of $\vec k$ and thus can be set to a constant 
$\Lambda $:
\begin{equation}
[2\epsilon_{k}-E]g(\vec k)= V\int_{k_F}^{k_F+k_D} g(\vec k')d^3k'=\Lambda
\end{equation}
giving
\begin{equation} 
V\int_{\epsilon_F}^{\epsilon_F+\hbar
  \omega_D}{\frac{N(\epsilon)d\epsilon}{[2\epsilon-E]}} = 1
\end{equation} 
where $N(\epsilon)$ is the density of the states of the (Bloch) fermions at the energy $\epsilon$, and 
$\epsilon_F$ is the Fermi energy. $N(\epsilon)\approx N(\epsilon_F)$ as 
$\hbar\omega_D<<\epsilon_F$, and the above integral gives us
\begin{equation} 
\ln{\frac{\Delta +\hbar\omega_D}{\Delta}}=\frac{2}{N(\epsilon_F)V}
\end{equation} 
where, $\Delta=2\epsilon_F-E$ is the energy gap. For weak coupling, $N(\epsilon_F)V<<1$ and we 
get
\begin{equation} 
\Delta\approx2\hbar\omega_D e^{\frac{2}{N(\epsilon_F)V}}
\end{equation} 
This energy gap $\Delta$ is related to the energy difference between normal and superconducting 
ground states:
\begin{equation} 
\langle E_n \rangle - \langle E_s \rangle=\frac{1}{2}N(\epsilon_F)\Delta^2
\end{equation} 
The full many-electron ground state is a coherent superposition of
such paired states.

\section{Ground state for the case $\Omega \neq 0$}
In Moyal spacetime, functions compose through star-product as 
\begin{equation}
(f \star g)(x) \equiv m_0({\cal F}^{-1}f\otimes g)(x)\equiv
 f(x) e^{\frac{i}{2}\Omega \epsilon_{ij} \overleftarrow{\partial}_{i} 
\overrightarrow{\partial}_{j}} g(x), 
\end{equation}
where,
\begin{equation}
{\cal F}=e^{\frac{i}{2}\Omega \epsilon_{ij} \hat p_i \otimes \hat p_j}
\label{twst}
\end{equation}
is the twist operator and $m_0$ is the point-wise multiplication
map. Note that here we are considering a simplified form of the
noncommutative parameter, so that it is essentially planer with $\hat
z$ commuting with both $\hat x$ and $\hat y$. This implies that
$\theta_{ij}=\Omega \epsilon_{ij}$ (for $i,j=1,2$) and $\theta_{13}=\theta_{23}=0$. 
The implementation of rotational symmetry in
the twisted framework now requires that the transformation properties
of multiparticle states (in contrast to the single particle state)
should also be deformed. This is captured in the deformed coproduct
(see \cite{bmpv,bgmpqv} for the detailed mathematical descriptions)
which in turn implies that the exchange operator $\tau_0$, defined as
$\tau_0 : \psi \otimes \phi \rightarrow \phi \otimes \psi$, should
also be deformed as $\tau_{\theta}={\cal F}_{\theta}^{-1}\tau_0{\cal
  F}_{\theta}$. One is thus forced to introduce the concept of twisted
fermion/boson by ``twist''-(anti) symmetrizing by using the projector
$P_{\theta}=\frac{1}{2}(1\pm \tau_{\theta})$ . It then turns out that in
noncommutative space, the fermionic creation and annihilation
operators in momentum basis satisfy the twisted anti-commutation
relations \cite{bmpv,bgmpqv}
\begin{eqnarray}
a_{\vec k,\sigma}a_{\vec k',\sigma'} &+& e^{\vec k \wedge \vec k'}a_{\vec k',\sigma'}
a_{\vec k,\sigma} = 0, \nonumber \\
a^{\dagger}_{\vec k,\sigma}a^{\dagger}_{\vec k',\sigma'} &+& e^{i \vec k \wedge \vec k'}
a^{\dagger}_{\vec k',\sigma'} a^{\dagger}_{\vec k,\sigma} = 0, \nonumber \\
a_{\vec k,\sigma}a^{\dagger}_{\vec k',\sigma'} &+& e^{-i\vec k \wedge \vec k'}
a^{\dagger}_{\vec k',\sigma'}a_{\vec k,\sigma}= \delta(\vec k-\vec k')\delta_{\sigma\sigma'},
\label{ts2}
\end{eqnarray} 
where, $\vec k\wedge \vec k'=\Omega \epsilon_{ij} k_{i}k'_{j}$. The twisted oscillators 
$a_{\vec k,\sigma}$ are related to their commutative (or untwisted) counterparts 
$c_{\vec k,\sigma}$ introduced in \cite{bmpv} as
\begin{eqnarray}
 a_{\vec k,\sigma} &=& c_{\vec k,\sigma}e^{-\frac{i}{2} \vec k \wedge {\cal P}} \label{a-c}, \\
{\cal P}_i &=& \int d^3 k \, k_i a^{\dagger}_{\vec k,\sigma} a_{\vec k,\sigma} 
= \int d^3 k \, k_i c^{\dagger}_{\vec k,\sigma} c_{\vec k,\sigma}
\end{eqnarray} 
where ${\cal P}_i$ is the Fock space momentum operator. Although the
single-particle state created by $a^\dagger_{\vec k,\sigma}$ is the
same as that created by $c^\dagger_{\vec k,\sigma}$, this is no longer
true for multi-particle states: in fact there is no observable that
connects a twisted multi-particle state to an untwisted multi-particle
state, as these belong to inequivalent superselection sectors \cite{bgmpqv}.

\subsection{Interactions}
We will restrict our attention to the case where $v(\vec x,\vec y)$ remains same in noncommutative 
case. This is a not unreasonable, as the potential $v$ is a function of different spatial points 
$\vec x$ and $\vec y$. The $\hat H^{int}_{\Omega}$ in the noncommutative space is thus 
\begin{equation} 
\hat H^{int}_{\Omega} = \frac{1}{2}\sum_{\sigma , \sigma'}\int d^3x \,
\psi_{\sigma}^{\dagger}(\vec x)\star
\Bigl[\int d^3y \, \psi_{\sigma'}^{\dagger}(\vec y) \star v(\vec x,\vec y)\star
\psi_{\sigma'}(\vec y) \Bigr]\star \psi_{\sigma}(\vec x) 
\end{equation} 
Writing $\psi_{\sigma}(\vec x)$ as
\begin{eqnarray}
\psi_{\sigma}(\vec x) = \int d^3k \,e^{-i\vec k.\vec x}a_{\vec k,\sigma}
\end{eqnarray}
where, $a_{\vec k, \sigma}, a^{\dagger}_{\vec k, \sigma} $ satisfy
the twisted commutation relations (\ref{ts2}), we can write $\hat H^{int}_{\Omega}$ as
\begin{eqnarray}
\hat H^{int}_{\Omega} = \frac{1}{2}\sum_{\sigma,\sigma'}\int d^3p\,d^3q\,d^3r\,d^3s
\langle pq|v|rs\rangle_{nc} \ a^{\dagger}_{\vec p,\sigma}
a^{\dagger}_{\vec q,\sigma'}a_{\vec r,\sigma'}a_{\vec s,\sigma} \, .
\end{eqnarray}
The matrix element $\langle pq|v|rs\rangle_{nc}$ is
\begin{eqnarray}
\langle pq|v|rs \rangle_{nc} &=& \int d^3x \,e^{i\vec p.\vec x}\star \Bigl[\int d^3y \,
e^{i\vec q.\vec y}\star v(\vec x, \vec y)\star e^{-i\vec r.\vec y} \Bigr]
\star e^{-i\vec s.\vec x} \, , \\
&=&e^{-\frac{i}{2}(\vec p\wedge \vec s + \vec q \wedge \vec r)}
\int \int e^{i\vec p.\vec x}e^{i\vec q.\vec y}v(\vec x,\vec y)e^{-i\vec r.\vec y}e^{-i\vec s.\vec x}  \nonumber \\
&=& e^{-\frac{i}{2}(\vec p\wedge \vec q + \vec q \wedge \vec r + \vec r \wedge \vec p)}
\tilde V(\vec r-\vec q) \delta_{(\vec r- \vec q),(\vec p -\vec s)} \label{matrixelt2}
\end{eqnarray}
The interaction Hamiltonian $\hat H^{int}_{\Omega}$ can be written in a 
manifestly twisted symmetrized form as
\begin{equation}
\hat H^{int}_{\Omega} = \frac{1}{2}\sum_{\sigma,\sigma'}\int d^3p\,d^3q\,d^3r\,d^3s\,
\Phi_{\Omega}(p,q,r,s) \ a^{\dagger}_{\vec p,\sigma}
a^{\dagger}_{\vec q,\sigma'}a_{\vec r,\sigma'}a_{\vec s,\sigma} \, ,
\end{equation}
where
\begin{eqnarray}
\Phi_{\Omega}(p,q,r,s) &\equiv& \bigl[\langle pq|v|rs\rangle_{nc} +  
e^{-i(\vec p \wedge \vec q + \vec r \wedge \vec s)} \langle qp|v|sr\rangle_{nc}\bigr] \quad {\rm and} \\
\Phi_{\Omega}(q,p,s,r) &=& e^{i(\vec p \wedge \vec q + \vec r \wedge \vec s)}\Phi_{\Omega}(p,q,r,s).
\end{eqnarray}
Using (\ref{matrixelt2}), $\hat H^{int}_{\Omega}$  takes the form
\begin{equation}
\hat H^{int}_{\Omega} = \frac{1}{2}\sum_{\sigma,\sigma'}\int d^3p\,d^3q\,d^3r \,
\tilde V_{\Omega}(\vec p, \vec q, \vec r) a^{\dagger}_{\vec p,\sigma}
a^{\dagger}_{\vec q,\sigma'}a_{\vec r,\sigma'}a_{(\vec p+\vec q -\vec r),\sigma}
\end{equation}
where $V_{\Omega}$ is given by,
\begin{eqnarray}
\tilde V_{\Omega} = e^{-\frac{i}{2}(\vec p \wedge \vec q + \vec q \wedge \vec r  +
  \vec r \wedge \vec p)}\tilde V(\vec r -\vec q)  
\label{27}
\end{eqnarray}
The free part of the Hamiltonian 
\begin{equation}
\hat H^0_{\Omega} = \int_{k_F} d^3k \, \epsilon(\vec k) a^{\dagger}_{\vec k, \sigma} 
a_{\vec k, \sigma} = \int_{k_F} d^3k \, \epsilon(\vec k) c^{\dagger}_{\vec k, \sigma} c_{\vec k, \sigma} 
= \hat H^0
\end{equation}
is the same as in the commutative case because of (\ref{a-c}). Therefore the full Hamiltonian is
\begin{equation}
\hat H_{\Omega}=\hat H^{0}_{\Omega} + \hat H^{int}_{\Omega}
\end{equation}

\subsection{Gap equation}
We wish to solve the eigenvalue equation
\begin{equation}
(\hat H^0_{\Omega} + \hat H^{int}_{\Omega})|\psi\rangle=E|\psi\rangle
\end{equation}
for $|\psi\rangle=\int d^3k\,g_\Omega (\vec k)|\psi_{\vec k}\rangle =
\int d^3k\,g_\Omega (\vec k) a_{\vec k, \frac{1}{2}}^\dagger a_{-\vec
  k, -\frac{1}{2}}^\dagger | F \rangle$ a paired 
state of twisted fermions. Note that the ansatz (3.18) has the similar form as 
that of its commutative counterpart (2.3) with deformed coefficient 
$g_{\Omega(\vec k)}$. This is because here we are interested in understanding the 
robustness of standard BCS type of superconductivity (in the regime of small 
Berry curvature), where one considers the 
even parity spin singlet states (particularly $l=0$ state {\it i.e.} s-wave 
superconductivity). Indeed, it has been shown explicitly in \cite{cond06} that 
higher pairing channels e.g. p-wave channel do not get activated in presence 
of small Berry curvature $\Omega$; the activation of higher pairing channels require the 
presence of large $\Omega$.  
 Furthermore, the twist operator (\ref{twst}) shows that 
it acts only on the configuration space when twist symmetrizing, leaving spin 
part untouched so that the spin singlet (or triplet) states can be constructed 
in the usual manner see \cite{bmpv, bgmpqv}. 

Because of the map (\ref{a-c}), the number operator in noncommutative case is the same as 
its commutative counterpart ($a^{\dagger}_{\vec k}a_{\vec k}=c^{\dagger}_{\vec k}c_{\vec k}$), giving 
us $\hat H^0_{\Omega}|\psi_{\vec k}\rangle=2\epsilon(\vec k)|\psi_{\vec k}\rangle$. But, since 
the operators $a_{\vec k}, a^{\dagger}_{\vec k}$ satisfy the twisted anti-commutation relations 
(\ref{ts2}), we get
\begin{eqnarray}
\hat H^{int}|\psi\rangle &=& \int d^3k\,g_\Omega(\vec k)\hat H^{int}|\psi_{\vec
  k}\rangle \nonumber \\ 
&=& \int d^3p\,d^3k \,V_{\Omega}(\vec p,\vec k)g_\Omega (\vec k)|\psi_{\vec p}\rangle
\end{eqnarray}
where, $V_{\Omega}(\vec p,\vec k)$ is given, in terms of $\tilde
V_{\Omega}$ (\ref {27}), by
\begin{eqnarray}
V_{\Omega}(\vec p, \vec k) &=& \frac{1}{2}\bigl[\tilde
  V_{\Omega}(\vec p,-\vec p, -\vec k) +
\tilde V_{\Omega}(-\vec p,\vec p,\vec k) \bigr]\\
&=& e^{-i\Omega \epsilon_{ij}p_i k_j}\tilde V(\vec p -\vec k)
\end{eqnarray}
In the noncommutative case the eigenvalue equation $\hat H_{\Omega}
|\psi\rangle=E|\psi\rangle$ gives
\begin{eqnarray}
[2\epsilon(\vec k)-E]g_\Omega (\vec k) = -\int d^3k' \,V_{\Omega}(\vec k,\vec k')g_\Omega(\vec k')
\label{Seqn3}
\end{eqnarray}
Assuming that $V(\vec p -\vec k)$ is constant $(-V)$ when $k_{F}\le k, k'\le k_F+\hbar \omega_D$ and
otherwise $0$, (\ref{Seqn3}) takes the form
\begin{eqnarray}
[2\epsilon(\vec k)-E]g_\Omega(\vec k)=V \int d^3k' \,g_\Omega(\vec k')e^{-i\Omega \epsilon_{ij} k_i k'_j}
\label{ncgapeqn}
\end{eqnarray}

\subsection{Solution of the gap equation}
It is easy to see that for modes of equal and opposite momenta, the deformed anti-commutation 
relations are 
\begin{equation} 
a^{\dagger}_{-\vec k, \sigma'}a^{\dagger }_{\vec k, \sigma}=
-e^{i \Omega \epsilon_{ij}(k_i)(-k_j)}  
a^{\dagger }_{\vec k,\sigma}a^{\dagger}_{-\vec k,\sigma'} =
-a^{\dagger }_{\vec k, \sigma}a^{\dagger}_{-\vec k, \sigma'} 
\end{equation} 
implying that the composite wave function of two twisted fermions with equal and opposite momenta 
is anti-symmetric like its commutative counterpart.  As the fermions are in a spin-singlet state, the spin 
part of the wave-function is anti-symmetric, thus forcing the momentum (or position) space wave 
function to be symmetric. This requires that $g(\vec k)=g(-\vec k)$. We can thus write the gap 
equation (\ref{ncgapeqn}) in the form
\begin{eqnarray}
[2\epsilon(\vec k)-E]g_\Omega(\vec k)=V \int d^3k' \, \cos(\Omega \epsilon_{ij}k_{i}k'_{j})g_\Omega(\vec k').
\label{ncgapeqn2}
\end{eqnarray}

We now solve the gap equation (\ref{ncgapeqn2}) perturbatively  in $\Omega$. To that end, let us 
write $g(\vec k)$ and $E$ in a series expansion as
\begin{eqnarray}
g_{\Omega}(\vec k) &=& g_0(\vec k) + \sum_{n=1}^{\infty}\Omega^n g_{n}(\vec k), \\ 
E &=& E_0 + \sum_{n=1}^{\infty}\Omega^nE_n.
\end{eqnarray}
Expanding $\cos(\Omega(k_1k'_2-k_2k'_1))$ in a Taylor series in $\Omega$ 
and equating coefficients of various powers of $\Omega$ on the both sides of (\ref{ncgapeqn2}) we 
get
\begin{eqnarray}
f_0(k)g_0(\vec k) &=& V \int d^3k' \,g_0(\vec k'),  \label{0order}\\
f_0(\vec k)g_1(\vec k)-E_1g_0(\vec k) &=& V \int d^3k' \,g_1(\vec k'), \label{1order} \\
f_0(\vec k)g_2(\vec k)-E_1g_1(\vec k)-E_2g_0(\vec k)&=&
\frac{V}{2}\Bigl[2\int g_2(\vec k')d^3k' - k^{2}_1\int
  k'^{2}_2g_0(\vec k')d^3k'  \nonumber \\ 
&-& k^{2}_2 \int k'^{2}_1g_0(\vec k')
+ 2 k_1 k_2 \int k'_1k'_2g_0(\vec k')d^3k' \Bigr] \label{2order}
\end{eqnarray} 
where $f_0(k)=\epsilon(\vec k)-E_0=\frac{{\vec k}^2}{2m}-E_0$.  From (\ref{0order}) we see that 
$g_0(\vec k)$ is the same as $g(\vec k)$ in the commutative case and is spherically symmetric. Using 
the spherical symmetry of $g_0$ we are able write (\ref{2order}) in a more simplified form as
\begin{eqnarray}
f_0(\vec k)g_2(\vec k)-E_1g_1(\vec k)-E_2g_0(\vec k)=V\int g_2(\vec k')d^3k'-
Vk^{2}_1\int k'^{2}_2g_0(\vec k')d^3k'
\end{eqnarray}   
Solving (\ref{0order},\ref{1order},\ref{2order}) we get
\begin{eqnarray}
E_1 &=& 0, \nonumber \\
 E_2 &=& V\frac{\beta^2}{\gamma}
\end{eqnarray}   
where
\begin{eqnarray}
\beta &=& \int \frac{k_1^{2}}{\epsilon(\vec k)-E_0}d^3k, \\
\gamma &=& \int_{k_F}^{k_F+\hbar\omega_D} \frac{1}{[2\epsilon(\vec k)-E_0]^2}d^3k \nonumber \\
&=& \int_{\epsilon_F}^{\epsilon_F + \hbar\omega_D} \frac{N(\epsilon)d\epsilon}{[2\epsilon-E_0]^2}  
\end{eqnarray}   
We can finally write the energy gap in for the $\Omega \neq 0$ case as 
\begin{eqnarray}
\Delta_{\Omega}=\Delta_0-\Omega^2 E_2
\end{eqnarray}   
where, $\Delta_0$ is the gap in the commutative case and is given by,
\begin{eqnarray}
\Delta_0 = \frac{k_{F}^2}{2m} - E_0\approx2\hbar\omega_D e^{\frac{2}{N(\epsilon_F)V}} 
\label{ncgap}
\end{eqnarray}   
This shows that to second order in $\Omega$, the energy gap {\it reduces} in the presence of noncommutativity. This is reminiscent of the presence of an external magnetic field \cite{sigr}: in either case time-reversal symmetry is broken (see 
\cite{sg} for time reversal symmetry breaking in noncommutative system and 
its impact on lifting of degeneracy).

\section{Conclusion}
In this paper we have investigated the effect of spatial noncommutativity (of Moyal type) in the 
Cooper-like problem of BCS superconductivity, and find that at least in the second order of the
noncommutative parameter there is an effective reduction in the gap energy. This is reminiscent of 
the behaviour of a superconductor in the presence of an external magnetic field, and it is tempting 
to suggest that noncommutativity provides a very simple model of this property. As has been pointed 
out in the introduction, that such an investigation is not of mere academic interest; 
genuine (but effective) noncommutativity can indeed be induced by the Berry curvature in 
a class of condensed matter systems. 
This analysis therefore gives a prominent and explicit role to the topological/geometrical 
properties of band structure and the consequent implications of this properties to the 
algebric structures in quantum field theory.   
 We would like to emphasize here that the effect of 
such 
noncommutativity stems from both the Moyal product and the twisted anti-commutation relations. It 
would be interesting to see if our results can be realized in experiments on ferromagnetic or GaAs 
crystals. 

Finally we would like to mention that the noncommutative structure given in eq.(1.1) or its simplified 
version eq.(1.2) in the absence of the external magnetic field ($B=0$) was 
obtained, in turn, from the semiclassical Lagrangian derived in Sundaram \& and Niu \cite{SN}
by using the Dirac bracket formalism appropriate for the second class constraints obtained from this Lagrangian. 
But this semiclassical Lagrangian itself was obtained within the approximations where terms involving higher moments 
of the wave packets {\it i.e.} those containing  higher order gradient terms in the perturbations were neglected. 
It is quite plausible that retaining those higher order terms and running through the entire constraint analysis again may indeed 
produce terms involving momentum space metric, which in turn involve second order derivatives like the Berry curvature. 
Our analysis is limited to the regime in which this derivation of ref.[5] holds.
In this context, we would like to mention that the momentum-space metric $g_{\alpha \beta}$ 
and Berry curvature corresponds to the real and imaginary  parts of a certain gauge invariant quantity as has been shown in 
\cite{MV}. If a small Berry curvature $\Omega$ is due to a small variation of the periodic part of the Bloch wave function 
w.r.t Bloch momentum, then 
correspondingly the metric $g_{\alpha \beta}$ will be nearly flat. As mentioned earlier, our analysis in this paper 
was essentially restricted to this domain. It will definitely be interesting to study the case of strong Berry curvature, 
where terms involving the momentum-space metric will not be ignorable any more. In addition, one would have to consider 
the effects of activating higher channel pairings.

\section*{Acknowledgement}
It is a pleasure to thank Jayanta Bhattacharjee for suggesting this
problem and valuable discussions, and Subroto Mukerjee for discussions
and for bringing \cite{Xiao} to our notice. The work of PB is
supported in part by a DST grant. BC acknowledges TWAS-UNESCO
associateship appointment at CBPF and CNPq for financial support. BC
would also like to thank Francesco Toppan, P.G.Castro and other
members of CBPF for their generous hospitality during his stay there,
when the part of the work was completed.

{}

\end{document}